\documentclass[epj]{svjour}
\usepackage{graphics}
\usepackage{graphicx}
\usepackage{latexsym}
\usepackage{amsmath,amssymb}
\usepackage{mathrsfs}
\usepackage{hhline}
\usepackage{dcolumn}
\usepackage{bm}
\usepackage[usenames,dvipsnames]{color}
\usepackage{supertabular}
\usepackage{cite} 
\usepackage[hyphens]{url}
\newcommand{\vect}[1]{\mbox{\boldmath $#1$}}

\newcommand{\eref}[1]{Eq.~\eqref{eq:#1}}
\newcommand{\figref}[1]{Fig.~\ref{fig:#1}}

\begin{document}
\title{Community detection in directed acyclic graphs}
\author{Leo Speidel\inst{1,}\inst{2,}\thanks{The two authors contributed equally to the work.} \and Taro Takaguchi\inst{2,}\inst{3,}$^{\rm a}$ \and Naoki Masuda\inst{4,}\thanks{\emph{\email{naoki.masuda@bristol.ac.uk}}}%
}                     
%
%
\institute{Department of Mathematical Informatics, The University of Tokyo, 7-3-1 Hongo, Bunkyo-ku, Tokyo 113-8656, Japan \and 
JST, ERATO, Kawarabayashi Large Graph Project, 2-1-2 Hitotsubashi, Chiyoda-ku, Tokyo 101-8430, Japan \and
National Institute of Informatics, 2-1-2 Hitotsubashi, Chiyoda-ku, Tokyo 101-8430, Japan \and
Department of Engineering Mathematics, University of Bristol, Woodland Road, Clifton, Bristol BS8 1UB, UK}
\date{Received: date / Revised version: date}
%
\abstract{
Some temporal networks, most notably citation networks, are naturally represented as directed acyclic graphs (DAGs). To detect communities in DAGs, we propose a modularity for DAGs by defining an appropriate null model (i.e., randomized network) respecting the order of nodes. We implement a spectral method to approximately maximize the proposed modularity measure and test the method on citation networks and other DAGs. We find that the attained values of the modularity for DAGs are similar for partitions that we obtain by maximizing the proposed modularity (designed for DAGs), the modularity for undirected networks and that for general directed networks. 
In other words, if we neglect the order imposed on nodes (and the direction of links) in a given DAG and maximize the conventional modularity measure, the obtained partition is close to the optimal one in the sense of the modularity for DAGs.
} 
\maketitle
\section{Introduction}
\label{intro}

Temporality and community structure are two common features present in various types of network data. Temporality of networks refers to nodes and links that vary over time.
For example, a friendship link between a given pair of individuals is not always used even if they are close friends of each other. The link would be only occasionally active as the two individuals meet and then separate. Temporally varying networks are collectively called temporal networks \cite{Holme2012%
%
%
}.
Community structure posits that nodes or links in networks can be classified into groups, called communities \cite{Fortunato2010}. Typically, a community is defined such that links are dense within a community and relatively sparse across different communities. Many networks in different domains have community structure.

The two features can be naturally combined into community detection in temporal networks, and several algorithms have been proposed to this aim. Examples include cost minimization when temporal non-smoothness is a part of the cost function
\cite{Chakrabarti2006,Tantipathananandh2007
}, optimization under temporal smoothness constraints \cite{Kawadia2012
}, methods based on the Potts model \cite{Ronhovde2011,
Fenn2009}, clique percolation \cite{Palla2007
}, decomposition of adjacency tensors
\cite{Gauvin2014
}, 
generalization of modularity for adjacency tensors \cite{Mucha2010,Bassett2013,Sarzynska2015}, link clustering \cite{Pandit2011},  and the minimum description length principle \cite{Sun2007}.

Related to temporal networks is the concept of directed
acyclic graph (DAG). DAGs are directed networks without directed cycles. In DAGs, nodes can be positioned within a layer structure such that links only emanate from a node in a higher layer to a node in a lower layer (Fig.~\ref{fig:schematic_ordered}). DAGs have been common as a tool for statistical inference for decades \cite{Pearl1988,Jordan2004%
%
%
}.
%
%
Equally importantly, we find various instances of DAGs in the real world such as some food webs \cite{Allesina2009}, some dominance hierarchy networks \cite{Shimoji2014,McDonald2012},
citation networks \cite{Price1965,Leicht2007,Karrer2009a,Karrer2009b}, family trees~\cite{Ott1999}, and phylogenetic networks \cite{Moret2004}.
%

Temporal networks can be mapped to DAGs in at least two ways.
First, citation networks, a type of temporal network, can be naturally mapped to DAGs. A citation network is a directed network in which a node represents an article such as a scientific paper, patent, or court decision, depending on the network, and a link is directed from the citing to the cited nodes \cite{Radicchi2012}. It is temporal in the sense that it grows over time due to the addition of new nodes and links \cite{Holme2012}.
%
%
In principle, the links are directed backwards in time because newer nodes can cite older nodes but not vice versa, which makes the network a DAG.
There may be links contradicting the arrow of time in real data sets, such as mutual citations~\cite{Radicchi2012}, which would violate the definition of a DAG. However, these links are relatively few (see section~\ref{sec:citation_net} for exemplar numbers).
 
Second, a family of temporal networks can be mapped to DAGs, as schematically shown in Fig.~\ref{fig:schematic_temporal}.
Consider a sequence of adjacency matrices indexed by time, i.e., $(A^{(t)})_{t=1,\ldots,T}$, where~$t$ is discrete time,~$(A^{(t)})_{ij}=1$ if~$i$ and~$j$ are adjacent to each other in the $t$th snapshot,
 and $(A^{(t)})_{ij}=0$ otherwise. By definition, $(A^{(t)})_{ij}=1$ implies that $i$ and $j$ contacted each other at some (continuous) time contained in the time interval $[t, t+1)$.
Such a representation of temporal networks as sequences of matrices can be induced by the temporal resolution of the recording device or by the aggregation of continuous-time temporal network processes over a finite time window to create a snapshot \cite{Holme2012}.
In Fig.~\ref{fig:schematic_temporal}, each node $i$ is duplicated $T$ times, and each duplicated node is labelled
$(i,t)$, where $1\le t\le T$. Therefore, there are $NT$ nodes in total in the representation shown in Fig.~\ref{fig:schematic_temporal}, where $N$ is the number of nodes in the original temporal network. We draw a link from~$(i,t)$ to~$(i,t+1)$ for every node $i$ ($1\le i\le N$) and $1\le t<T$. We also draw a link between two nodes in subsequent layers if they are connected in the corresponding time interval. In other words, we draw a link from~$(j,t)$ to~$(i,t+1)$ if $(A^{(t)})_{ij} = 1$. In this way, a temporal network given as a sequence of adjacency matrices is uniquely mapped to a DAG in which links only span between two subsequent layers. This type of representation and its variants have been used in the analysis of temporal networks
\cite{Pfitzner2013,Kempe2000,Kostakos2009}.

Community detection methods for temporal networks mentioned earlier in the present section are designed for the latter type of representation. However, to the best of our knowledge, community detection methods explicitly designed for the former type of temporal networks, or more generally DAGs, have not been proposed so far. 
Community structure in the former type of temporal networks has been studied using methods such as agglomerative hierarchical clustering \cite{Hopcroft2004}, conventional undirected and directed modularity~\cite{Chen2010,Leicht2007}, and the Infomap method~\cite{Rosvall2010}. These detection methods are designed for static networks. They neither incorporate the acyclic nature of citation networks nor temporal information such as the publication dates of articles. In these studies, temporal dynamics were analyzed once communities were obtained by the static methods. Therefore, if we apply existing methods, we have to map a temporal network of the former type to a static network by discarding the temporal information or to a snapshot representation. Both types of mapping
%
%
seem to be suboptimal given the natural representation of the original network as a DAG.

In the present study, we develop a community detection method which exploits the intrinsic temporality in the former type of temporal networks.
Technically, we propose a community detection method for general DAGs.
We do so by developing a modularity measure and a maximization method for DAGs. A key to this development is the observation that
the choice of a null model network characterizes communities to be detected. Null models randomize links while preserving some properties of the original network. Communities obtained by modularity maximization therefore are structures that are statistically surprising relative to the null model. Depending on the class of networks, specific null models have been proposed. Examples include networks without multilinks and selfloops \cite{Massen2005}, weighted networks \cite{Newman2004b},
%
%
directed networks \cite{Leicht2008,Kim2010}, multi-partite networks \cite{Barber2007,Guimera2007,Murata2009},
%
%
spatial networks \cite{Expert2011
}, networks with a similarity measure imposed between nodes \cite{Liu2014
}, networks directly formed by correlation matrices \cite{MacMahon2015}, and multilayer networks \cite{Bassett2013,Sarzynska2015,Mucha2010} in which temporal networks are included as a special case.
If we use an inappropriate null model to detect communities in a DAG, we may obtain a suboptimal result. In particular, all aforementioned methods do not respect the directed layer structure inherent in a DAG. With such a method, a modularity value might be large simply because a DAG is surprising as compared to a non-DAG null model. However, we are interested in modular structure that a given DAG may have whereas reference DAGs do not. Therefore, in the present study we propose a modularity maximization method that uses a null model for DAGs.

\section{DAG}\label{sec:def of DAG}

\begin{figure}
\centering
\resizebox{0.4\textwidth}{!}{
\includegraphics[width=0.5\hsize]{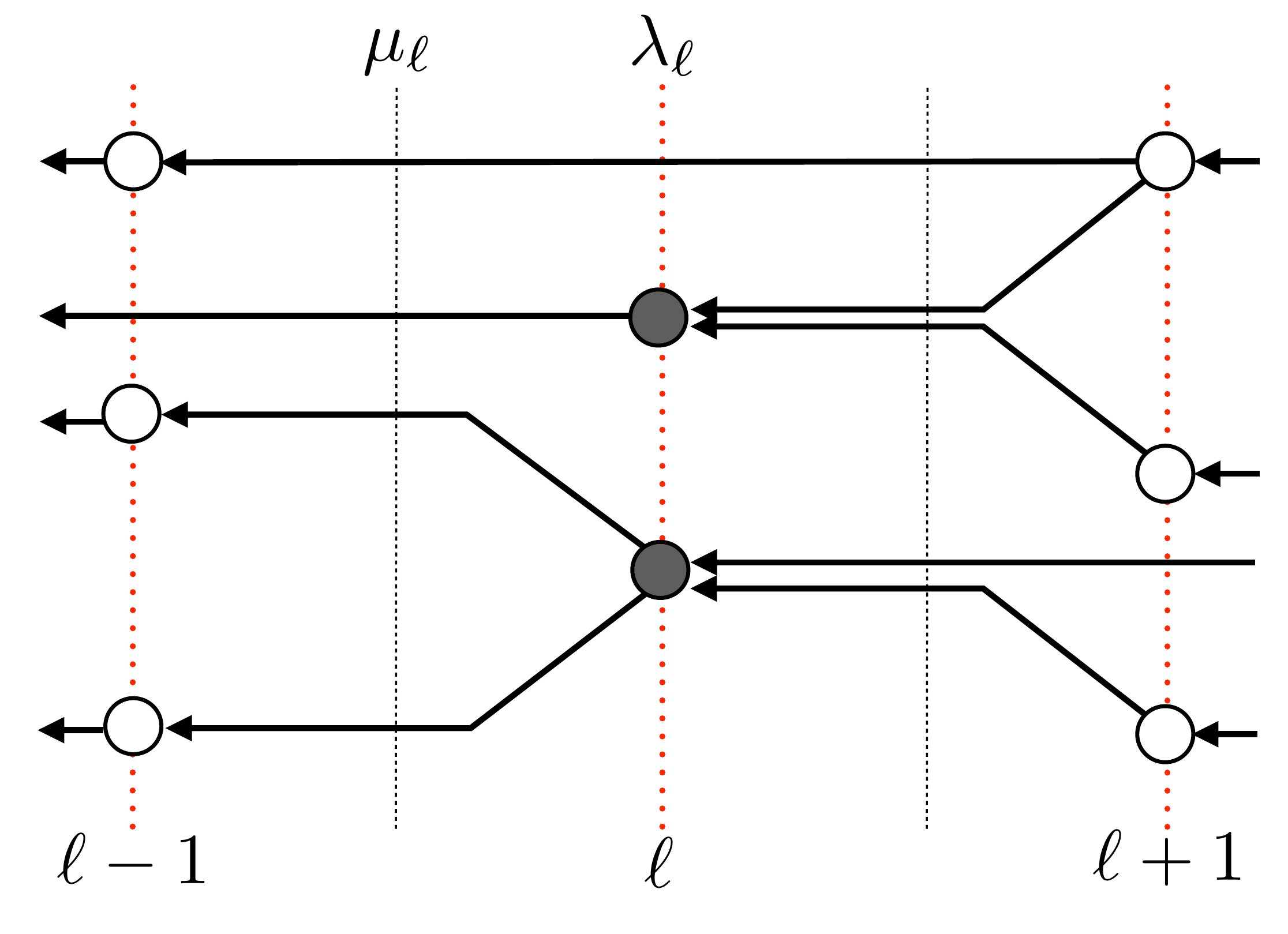}
}
\caption{
Schematic of a DAG magnified around layer~$\ell$. The indices~$\ell-1, \ell, \ell+1$ represent three subsequent layers from lower to upper.
The filled circles represent the nodes in layer~$\ell$. The quantities~$\mu_\ell$ and~$\lambda_\ell$ are given by the number of links passing through the corresponding vertical lines.}
\label{fig:schematic_ordered}
\vspace{10pt}
\centering
\resizebox{0.4\textwidth}{!}{
\includegraphics[width=0.5\hsize]{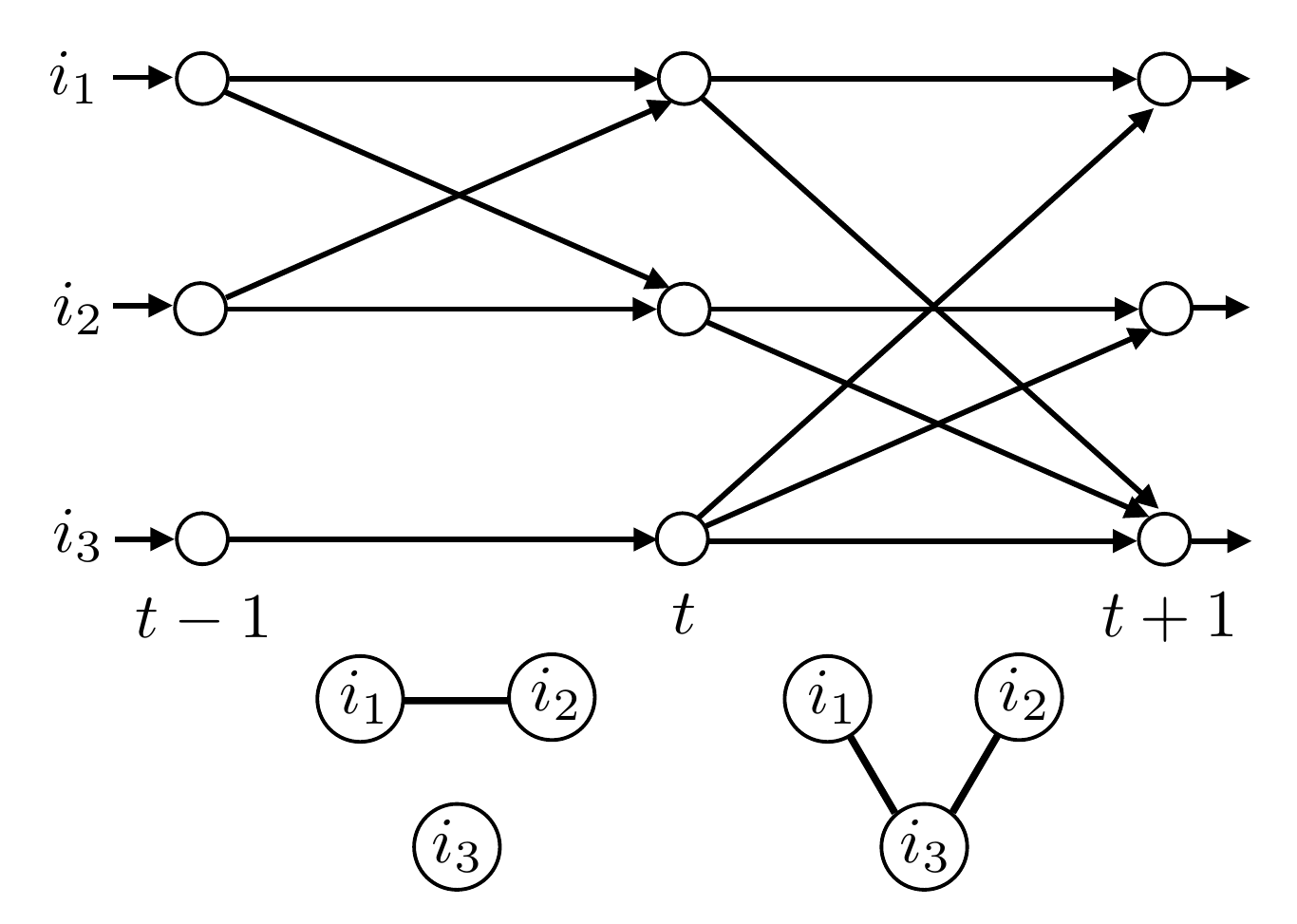}
}
\caption{Schematic of the DAG representation of a temporal network, in which the link configurations between the same set of nodes change over time. The lower row shows snapshots of the network taken at times in~$[t-1,t)$ and~$[t,t+1)$, respectively.}
\label{fig:schematic_temporal}
\end{figure}

Let us denote a directed network by~$G = (V, E)$ with~$|V| = N$ nodes and~$|E| = M$ directed links. We assume that~$G$ is a DAG. We also assume that $G$ has~$L$ layers, that node~$i$~$(1 \leq i \leq N)$ belongs to layer~$\ell_i$~$(1 \leq \ell_i \leq L)$, and that any directed link from $j$ to $i$ satisfies~$\ell_j > \ell_i$ (\figref{schematic_ordered}). In words, any link emanates from a node in the upper layer (i.e., larger layer number) to a node in the lower layer (i.e., smaller layer number). No directed 
link exists between nodes in the same layer. A layer may include multiple nodes.
For example, in Fig.~\ref{fig:schematic_ordered}, layers $\ell-1$, $\ell$, and $\ell+1$ contain three, two, and three nodes, respectively. No order is imposed between nodes in the same layer.

Our definition is motivated by empirical observations that it is often more appropriate not to impose an order in subsets of nodes. For example, in citation networks, articles published on the same date cannot cite each other unless the two papers coordinate their citations \cite{Radicchi2012,Leicht2007,Karrer2009a,Karrer2009b}. In dominance hierarchy networks of animals, it may be difficult to rank individual animals (i.e., nodes) if they have similar physical strengths and no prior interaction, which would be the case in large colonies.

We can always give such a layering to a given DAG although layering is not unique in general \cite{Eades1993}. In the remainder of the present article, we simply refer to a DAG supplied with a proper layering $\{\ell_1, \ldots, \ell_N\}$ as a DAG.


\section{Modularity}

Communities detected through modularity maximization depend on the null network model that serves as a baseline for defining the modularity. In this section, we first briefly review modularity measures for undirected and directed networks allowing for cycles. Then, we formulate modularity for DAGs by proposing a new null model tailored to the case of DAGs.

\subsection{Modularity for undirected networks}

The modularity for undirected networks is defined by
\begin{equation}
Q^{\rm und} \equiv \frac{1}{2M} \sum_{i,j=1}^N \left( A_{ij} - \frac{k_ik_j}{2M} \right) \delta_{c_i, c_j},  
\label{eq:Q_undirected}
\end{equation}
where~$A_{ij}$ is the~$(i, j)$ element of the adjacency matrix, and~$k_i=\sum_{j=1}^N A_{ij}$ is the degree of node~$i$ \cite{Newman2004,Fortunato2010}. The adjacency matrix is defined by~$A_{ij}=1$ if nodes $i$ and $j$ are adjacent by a link and $A_{ij}=0$ otherwise. It is symmetric (i.e., $A_{ij}=A_{ji}$) for undirected networks. Function~$\delta$ is the Kronecker delta, and~$c_i$ represents the index of the community to which node~$i$ belongs.
%
%
In Eq.~\eqref{eq:Q_undirected}, the frequency of links within communities, corresponding to
$A_{ij}\delta_{c_i, c_j}$ in \eref{Q_undirected}, is compared against that for the null model, corresponding to 
$k_ik_j\delta_{c_i, c_j}/2M$. Under the null model, the expected number of links between nodes $i$ and $j$ is equal to 
$P_{ij}^{\rm und}=k_ik_j/2M$. In other words, the null model is the configuration model in which each node has the same degree as in the original network.
%

\subsection{Modularity for directed networks}

The modularity for directed networks is defined by
\begin{equation}
Q^{\rm dir} \equiv \frac{1}{M} \sum_{i,j=1}^N \left( A_{ij} - \frac{k^{\rm in}_i k^{\rm out}_j}{M} \right) \delta_{c_i, c_j},  
\label{eq:Q_directed}
\end{equation}
where $k^{\rm in}_i$ and $k^{\rm out}_j$ are the in-degree of node~$i$ and the out-degree of node~$j$, respectively \cite{Arenas2007,
Leicht2008}. 
In general, $A$ is allowed to be asymmetric, and $A_{ij} = 1$ if there is a link from node~$j$ to~$i$ and~$A_{ij} = 0$ otherwise.
In~$Q^{\rm dir}$, the number of intra-community links is compared against the null model in which the expected number of links emanating from node $j$ to $i$ is equal to $P_{ij}^{\rm dir}=k_i^{\rm in}k_j^{\rm out}/M$.
In other words, the null model is the directed variant of the configuration model in which each node has the same in- and out-degrees as in the original network.
%

\subsection{Modularity for DAGs}

Although it is possible to apply $Q^{\rm und}$ and $Q^{\rm dir}$ to detect communities in DAGs, neither of them incorporates the partial order imposed on nodes in DAGs. Therefore, we propose a null model for DAGs by generalizing the random DAG model proposed in Refs.~\cite{Karrer2009a,Karrer2009b} (for other null models of DAGs, see Ref.~\cite{Goni2010}).
In the random DAG model, links are randomized while preserving the in- and out-degree of each node and the order of nodes as specified by $\ell_i$. The difference between the present definition and that in Refs.~\cite{Karrer2009a,Karrer2009b} is that the
former allows a layer to contain multiple nodes, whereas
each node constitutes a single layer in the latter case.

Following the derivation in Ref.~\cite{Karrer2009a,Karrer2009b}, we calculate the expected number of links $P_{ij}^{\rm dag}$ from node $j$ to node $i$. We start by defining
\begin{align}
\label{eq:mu_and_lambda}
\mu_\ell &\equiv \sum_{i; 1 \leq \ell_i < \ell} k_i^{\rm in} - \sum_{i; 1 \leq \ell_i < \ell} k_i^{\rm out},\\
\lambda_\ell &\equiv \sum_{i; 1 \leq \ell_i < \ell} k_i^{\rm in} - \sum_{i; 1 \leq \ell_i \leq \ell} k_i^{\rm out} = \mu_\ell - \sum_{i; \ell_i = \ell} k_i^{\rm out},
\end{align}
where $1 \leq \ell \leq L$.
In words,~$\mu_\ell$ is equal to the number of links from nodes in layers~$\left\{ \ell, \ell+1, \ldots, L \right\}$ to nodes in layers~$\left\{ 1, 2, \ldots, \ell -1 \right\}$, and $\lambda_\ell$ is equal to the number of links from nodes in layers~$\left\{ \ell+1, \ell+2, \ldots, L \right\}$ to nodes in layers~$\left\{ 1, 2, \ldots, \ell-1 \right\}$. Graphically,~$\lambda_\ell$ and~$\mu_\ell$ are equal to the number of links crossing a section at $\ell$ (dotted line in Fig.~\ref{fig:schematic_ordered}) and one before layer~$\ell$ (dashed line in Fig.~\ref{fig:schematic_ordered}), respectively. In the example shown in Fig.~\ref{fig:schematic_ordered},~$\lambda_{\ell} = 1$ and~$\mu_{\ell} = 4$. 

The in-degree of node $i$, $k_i^{\rm in}$, can be visualized as $k_i^{\rm in}$ in-stubs attached to node $i$, where a stub is a half-link. Likewise for the out-degree. Joining an in-stub and an out-stub yields a link.
We calculate the expected number for an in-stub attached to node~$i$ to connect to node~$j$, which we denote by~$p_{ij}$. Let us first consider the case~$\ell_j=\ell_i+1$. In this case, we obtain
\begin{equation}
\label{eq:Pord-P[connect]}
p_{ij} = \frac{k_j^{\rm{out}}}{\mu_{\ell_i+1}}
\end{equation}
because $j$ possesses $k_j^{\rm out}$ out-stubs and there are $\mu_{\ell_i+1}$ out-stubs from which an in-stub attached to
node $i$ can choose.
The probability that an in-stub of node~$i$ does not connect to a node in layer~$\ell_i+1$ is given by
\begin{equation}
\label{eq:Pord-P[miss]}
1-\sum_{j; \ell_j=\ell_i+1} p_{ij} = 1-\frac{\sum_{j; \ell_j = \ell_i+1} k_j^{\rm out}}{\mu_{\ell_i+1}} = \frac{\lambda_{\ell_i+1}}{\mu_{\ell_i+1}}.
\end{equation}
Now we assume in general that node~$j$ belongs to any layer~$\ell_j > \ell_i$. Then, an in-stub of node~$i$ can connect to node~$j$ only if it does not connect to a node in layers~$\ell_{i} < \ell < \ell_j$. By iteratively applying the argument leading to Eq.~\eqref{eq:Pord-P[miss]}, we find that the probability that an in-stub of node~$i$ does not to connect to any node in layers~$\ell_{i} < \ell < \ell_j$ is given by
\begin{equation}
\label{eq:Pord-P[miss_in_between]}
\prod_{\ell = \ell_i + 1}^{\ell_j-1} \frac{\lambda_{\ell_j}}{\mu_{\ell_j}}.
\end{equation} 
Provided that the in-stub of node~$i$ does not connect to a node in layers~$\ell_{i} < \ell < \ell_j$, the probability that it connects to node~$j$ is given by
\begin{equation}
\label{eq:Pord-P[connect_to_j]}
\frac{k_j^{\rm{out}}}{\mu_{\ell_j}}.
\end{equation}
Using Eqs.~\eqref{eq:Pord-P[miss_in_between]} and~\eqref{eq:Pord-P[connect_to_j]}, we obtain
\begin{equation}
p_{ij} = k_j^{\rm{out}} \frac{\prod_{\ell = \ell_i + 1}^{\ell_j-1} \lambda_{\ell_j}}{\prod_{\ell = \ell_i + 1}^{\ell_j}  \mu_{\ell_j}}.
\end{equation}

Because node $i$ has $k_i^{\rm{in}}$ in-stubs, the expected number of links from $j$ to $i$ is given by
\begin{equation}
P_{ij}^{\rm dag} = k_i^{\rm{in}} k_j^{\rm{out}} \frac{\prod_{\ell = \ell_i + 1}^{\ell_j-1} \lambda_{\ell_j}}{\prod_{\ell = \ell_i + 1}^{\ell_j}  \mu_{\ell_j}}.
\end{equation}
Using~$P_{ij}^{\rm dag}$, we define the modularity for DAGs, denoted by~$Q^{\rm dag}$, by
\begin{equation}
Q^{\rm dag} \equiv \frac{1}{M} \sum_{i,j} \left( A_{ij} - P_{ij}^{\rm dag}\right) \delta_{c_i, c_j},  
\label{eq:Q_ordered}
\end{equation}
where~$P_{ij}^{\rm dag}$ is given by
\begin{align}
P_{ij}^{\rm dag} \equiv
\begin{cases}
0 & (\ell_i \geq \ell_j),\\
k^{\rm in}_i k^{\rm out}_j \dfrac{\prod_{\ell = \ell_i + 1}^{\ell_j -1} \lambda_\ell}{\prod_{\ell = \ell_i + 1}^{\ell_j} \mu_\ell} & (\ell_i < \ell_j).
\end{cases}
\end{align}
%




\section{Spectral methods for modularity maximization}

Because maximization of modularity is a combinationally hard problem,
many approximate algorithms have been proposed for maximizing $Q^{\rm und}$ \cite{Fortunato2010}, whose variants are also used to maximize $Q^{\rm dir}$ \cite{Leicht2008,Fortunato2010,Malliaros2013}.
Here we adapt the spectral method \cite{Newman2006,Newman2006a} to the case of
$Q^{\rm dag}$.

\subsection{Spectral method for undirected networks}

Before describing the spectral method for~$Q^{\rm dag}$, we review the method for~$Q^{\rm und}$ \cite{Newman2006,Newman2006a}.
The spectral method realizes community detection by iteratively bipartitioning a tentative community.

We initially partition the entire network into two groups of nodes.
To this end, we use the fact that $Q^{\rm und}$ is written in matrix form as
\begin{equation}
Q^{\rm und} = \frac{1}{4M} \vect{s}^\top B^{\rm und} \vect{s},
\label{eq:quad_form_Q}
\end{equation}
where $\top$ denotes the transposition, $\vect{s} = (s_1\; \cdots\; s_N)^{\top}$, $s_i\in \{ -1, 1\}$, and~$B^{\rm und}$ is an $N \times N$ symmetric matrix whose elements are given by~$B^{\rm und}_{ij} \equiv A_{ij} - (k_i k_j / 2M)$.
Node~$i$ is classified to the first and second groups if $s_i=-1$ and $s_i = 1$, respectively.

Finding the vector~$\vect{s}$ that maximizes~$Q^{\rm und}$ is an NP-complete problem~\cite{Brandes2008}. A commonly applied heuristics is to relax elements of~$\vect{s}$ to take continuous values and impose the normalization constraint~$\vect{s}^\top\vect{s}=N$. Then, $Q^{\rm und}$ is maximized when $\vect{s}$ is the eigenvector associated with the largest eigenvalue of $B^{\rm und}$, which we denote by~$\vect{u}$.
We carry out bipartition by setting $s_i = {\rm sgn} (u_i)$~$(1 \leq i \leq N)$.

The spectral method takes advantage of the fact that we can rapidly calculate $\vect{u}$ using the power method. In fact, the power method requires calculation of the product $B^{\rm und}\vect{x}$ for changing $\vect{x}$. Although this computation may look computationally costly because $B^{\rm und}$ is a dense matrix, $B^{\rm und}\vect{x}$ can be expressed as
\begin{equation}
B^{\rm und} \vect{x} = A \vect{x} - \frac{\vect{k} (\vect{k}^\top \vect{x})}{2M},
\label{eq:Bx_und}
\end{equation}
where~$\vect{k} \equiv (k_1, k_2, \ldots, k_N)^\top$~\cite{Newman2006,Newman2006a}.
The inner product $\vect{k}^\top \vect{x}$ is calculated in time~$O(N)$, while the calculation of~$A \vect{x}$ requires time~$O(N+M)$. For sparse graphs, $M = O(N)$ such that $B^{\rm und}\vect{x}$ can be calculated in $O(N)$ time, accelerating the power method.

Once the network is partitioned into two tentative communities, we repeat selecting and bipartitioning one tentative community until $Q^{\rm und}$ stops increasing. To decide whether to approve a proposed bipartition of a tentative community
$C$ into two groups of nodes $C_1$ and~$C_2$, one needs to calculate the change in modularity by the partitioning, which is given by
\begin{align} 
\label{eq:delta_Qund}
\Delta Q^{\rm und}
= \frac{1}{2M} \left[ \sum_{i \in C_1, j \in C_2} \left(B^{\rm und}_{ij} + B^{\rm und}_{ji}\right) - \sum_{i,j \in C} B^{\rm und}_{ij} \right].
\end{align}
In matrix form, $\Delta Q^{\rm und}$ is written as
\begin{equation}
\Delta Q^{\rm und} = \frac{1}{4M} \tilde{\vect{s}}^\top \tilde{B}^{\rm und} \tilde{\vect{s}},
\label{eq:quad_form_Qund}
\end{equation}
where~$\tilde{\vect{s}}$ is a column vector with length~$|C|$, and $\tilde{B}^{\rm und}$ is a $|C| \times |C|$ symmetric matrix whose elements are defined by~$\tilde{B}^{\rm und}_{ij} \equiv B^{\rm und}_{ij} - \delta_{ij}\sum_{k \in C} B^{\rm und}_{ik}$
\cite{Newman2006}. When $C$ is the entire network, $\Delta Q^{\rm und}$, $\tilde{B}^{\rm und}$, and $\tilde{\vect{s}}$ are equal to $Q^{\rm und}$, $B^{\rm und}$, and $\vect{s}$, respectively.
Under the constraint $\tilde{\vect{s}}^\top\tilde{\vect{s}}=|C|$, Eq.~\eqref{eq:quad_form_Qund} is maximized if~$\tilde{\vect{s}}$ is the eigenvector associated with the largest eigenvalue of~$\tilde{B}^{\rm und}$, which we denote by $\tilde{\vect{u}}$.
Finally, we bipartition $C$ by setting $\tilde{s}_i = {\rm sgn} (\tilde{u}_i)$ $(i \in C)$.

\subsection{Spectral method for directed networks}

The spectral method for directed networks was implemented in Ref.~\cite{Leicht2008}. Here we briefly recapitulate their arguments. We can rewrite the modularity for directed network as $Q^{\rm dir} = (1 / 4M) \vect{s}^\top \left[B^{\rm dir} + (B^{\rm dir})^\top\right] \vect{s}$ with $(B^{\rm dir})_{ij} \equiv A_{ij} - (k_i^{\rm in} k_j^{\rm out}) / M$. Because $B^{\rm dir}+(B^{\rm dir})^\top$ is a symmetric matrix, we can maximize $Q^{\rm dir}$ by replacing $B^{\rm und}$ in Eq.~\eqref{eq:Bx_und} by $B^{\rm dir}+(B^{\rm dir})^{\top}$ and applying the same spectral method as that for $Q^{\rm und}$.

To calculate the change in modularity $\Delta Q^{\rm dir}$ caused by bipartitioning, we replace $B^{\rm und}$ by $B^{\rm dir}+(B^{\rm dir})^\top$ in Eqs.~\eqref{eq:delta_Qund} and \eqref{eq:quad_form_Qund}. Then, we maximize $\Delta Q^{\rm dir}$ by following the same steps as for maximizing $\Delta Q^{\rm und}$.

\subsection{Spectral method for DAGs}

The application of the spectral method to $Q^{\rm dag}$ is straightforward. For $Q^{\rm dag}$, a single step in the power method can be accelerated as in the case of $Q^{\rm und}$ and $Q^{\rm dir}$ as follows. 

First, we write
\begin{equation}
Q^{\rm dag} = \frac{1}{4M} \vect{s}^\top \left[ B^{\rm dag} + (B^{\rm dag})^\top \right] \vect{s},
\end{equation}
where~$B^{\rm dag}_{ij} \equiv A_{ij} - P^{\rm dag}_{ij}$.
To find a decomposition of $\left[B^{\rm dag}+(B^{\rm dag})^{\top}\right]\vect{x}$ similar to \eref{Bx_und},
we write
\begin{equation}
P_{ij}^{\rm dag} = k^{\rm in}_i k^{\rm out}_j  f(\ell_i, \ell_j),
\end{equation}
where
\begin{align}
f(\ell_i, \ell_j) \equiv
\begin{cases}
0 & (\ell_i \geq \ell_j),\\
\dfrac{\prod_{\ell = \ell_i + 1}^{\ell_j -1} \lambda_\ell}{\prod_{\ell = \ell_i + 1}^{\ell_j} \mu_\ell} & (\ell_i < \ell_j).
\end{cases}
\label{eq:f(ell_i,ell_j)}
\end{align} 
As discussed in detail in Ref.~\cite{Karrer2009b}, Eq.~\eqref{eq:f(ell_i,ell_j)} when $\ell_i < \ell_j$ is identical to
\begin{equation}
f(\ell_i, \ell_j) = 
\begin{cases}
\frac{f(\ell_i, \ell^{\rm up}(j)) f(\ell^{\rm low}(i), \ell_j)}{f(\ell^{\rm low}(i), \ell^{\rm up}(j))} & (\lambda_{\ell_i+1}, \ldots, \lambda_{\ell_j-1} \neq 0),\\
0 & (\text{otherwise}),
\end{cases}
\label{eq:f(ell_i,ell_j) new expression}
\end{equation}
where, $\ell^{\rm low}(i)$ is the largest value of $\ell (\le  \ell_i)$ such that $\lambda_{\ell}=0$, and
$\ell^{\rm up}(j)$ is the smallest value of $\ell (\ge \ell_j)$ such that  $\lambda_{\ell}=0$.
If there is no $\ell (\le \ell_i)$ or $\ell (\ge \ell_j)$ such that $\lambda_{\ell}=0$, we obtain $\ell^{\rm low}(i)=1$
and $\ell^{\rm up}(j) = L$, respectively. In words, $f(\ell_i, \ell_j) > 0$ if and only if for every layer between $\ell_i$ and $\ell_j$, we find at least one link penetrating the layer.
Using Eq.~\eqref{eq:f(ell_i,ell_j) new expression}, we decompose~$P_{ij}^{\rm dag}$ in the case $\ell_i < \ell_j$ and 
$\lambda_{\ell_i+1}, \ldots, \lambda_{\ell_j-1} \neq 0$ as
\begin{equation}
\label{eq:P_ijord = f..}
P_{ij}^{\rm dag} = \frac{f(\ell_i, \ell^{\rm up}(j))}{f(\ell^{\rm low}(i), \ell^{\rm up}(j))}k_i^{\rm in} \cdot f(\ell^{\rm low}(i), \ell_j) k_j^{\rm out}.
\end{equation}
It should be noted that $P_{ij}^{\rm dag}=0$ otherwise.
Because $\lambda_{\ell_i+1}, \ldots, \lambda_{\ell_j-1} \neq 0$ if and only if $\ell^{\rm low}(i) = \ell^{\rm low}(j)$ and $\ell^{\rm up}(i) = \ell^{\rm up}(j)$, we can rewrite Eq.~\eqref{eq:P_ijord = f..} as
\begin{equation}
\label{eq:P_ijord = f..2}
P_{ij}^{\rm dag} = \frac{f(\ell_i, \ell^{\rm up}(i))}{f(\ell^{\rm low}(i), \ell^{\rm up}(i))}k_i^{\rm in} \cdot f(\ell^{\rm low}(j), \ell_j) k_j^{\rm out}.
\end{equation}
For each layer~$\ell$, we define $\vect{\kappa}^{\rm in}(\ell) = (\kappa^{\rm in}_1(\ell)\; \cdots\; \kappa^{\rm in}_N(\ell))^{\top}$ and $\vect{\kappa}^{\rm out}(\ell) = (\kappa^{\rm out}_1(\ell)\; \cdots\; \kappa^{\rm out}_N(\ell))^{\top}$ by
\begin{align}
\label{eq:kappa_in}
\kappa_i^{\rm in}(\ell) =
\begin{cases}
0 & (\ell_i \neq \ell),\\
\dfrac{f(\ell_i, \ell^{\rm up}(i))}{f(\ell^{\rm low}(i), \ell^{\rm up}(i))}k_i^{\rm in} & (\ell_i = \ell),\\
\end{cases}
\end{align}
and
\begin{align}
\label{eq:kappa_out}
\kappa_i^{\rm out}(\ell) =
\begin{cases}
0 & (\ell < \ell^{\rm low}(i) \textrm{ or } \ell_i \leq \ell),\\
f(\ell^{\rm low}(i), \ell_i) k_i^{\rm out} & (\ell^{\rm low}(i) \le \ell < \ell_i),
\end{cases}
\end{align}
respectively. By combining Eqs.~\eqref{eq:P_ijord = f..2},~\eqref{eq:kappa_in} and~\eqref{eq:kappa_out}, we obtain
\begin{equation}
P_{ij}^{\rm dag} = \kappa_i^{\rm in}(\ell_i) \kappa_j^{\rm out}(\ell_i) = \sum_{\ell = 1}^L \kappa_i^{\rm in}(\ell) \kappa_j^{\rm out}(\ell).
\end{equation} 
Finally, we decompose~$B^{\rm dag} \vect{x}$ as
\begin{equation}
B^{\rm dag} \vect{x} = A \vect{x} - \sum_{\ell = 1}^{L} \vect{\kappa}^{\rm in}(\ell) \left( \vect{\kappa}^{\rm out}(\ell)^\top \vect{x} \right).
\label{eq:Bx_ord}
\end{equation}
Therefore, once~$\left\{ \vect{\kappa}^{\rm in}(\ell), \vect{\kappa}^{\rm out}(\ell) \right\}_{1 \leq \ell \leq L}$ have been calculated beforehand, the computation of~$B^{\rm dag} \vect{x}$ takes $O(LN)$ time, and likewise for $(B^{\rm dag})^{\top}\vect{x}$.

To implement repeated bipartitioning, we maximize the change in modularity, $\Delta Q^{\rm dag}$, which a biparition of a community causes. We calculate $\Delta Q^{\rm dag}$ by replacing $B^{\rm und}$ by $B^{\rm dag}+(B^{\rm dag})^\top$ in Eqs.~\eqref{eq:delta_Qund} and \eqref{eq:quad_form_Qund}. Then, we follow the same steps as for maximizing $\Delta Q^{\rm und}$. 

\section{Results}
\label{sec:results}

\begin{table}
\centering
\caption{Properties of the networks analyzed in section~\ref{sec:results}. The number of nodes, that of links, and that of layers are denoted by $N$, $M$, and $L$, respectively. These values refer to those of the largest weakly connected component of each network.}
\label{tab:data}
\begin{tabular}{lccc}
\hline\noalign{\smallskip}
	Data	&  $N$	 	&	 $M$	&	 $L$	\\
\noalign{\smallskip}\hline\noalign{\smallskip}
HepPh 	&	30,337	& 	344,578	& 	3,683\\
HepTh 	&	27,377	& 	351,033	&	4,139\\
\textit{E.~coli}	&	328		&	456		&	5	\\
C1 		&	20		& 	29		&	4 	\\
C2 		&	32		& 	55		&	7	\\
C3		&	48		&	134		&	8	\\
C4		&	70		&	158		&	7	\\
C6		&	64		&	137		&	8	\\
\noalign{\smallskip}\hline
\end{tabular}
\end{table}

In this section, we test the proposed spectral method for maximizing $Q^{\rm dag}$ for several DAG data. 
We test the method against the spectral method maximizing $Q^{\rm und}$ and $Q^{\rm dir}$ and the so-called Louvain method, which is a non-spectral
algorithm that heuristically maximizes $Q^{\rm und}$ \cite{Blondel2008}. We use the implementation of the Louvain method in the igraph package of R \cite{Csardi2006}. In our implementation of the spectral method, we add a fine tuning step after every bipartitioning in a similar fashion to Refs.~\cite{Newman2006,Newman2006a} (Appendix~\ref{app:fine tuning}). We do not apply this fine tuning to the Louvain method because this fine tuning is specialized to the spectral method. We also apply a postprocessing procedure after the spectral method or the Louvain method has terminated
(Appendix~\ref{app:postprocessing}). To apply the spectral method and the Louvain method to maximize $Q^{\rm und}$, and also to calculate $Q^{\rm und}$ for the final partition
%
%
obtained by various methods, we ignore link direction, or equivalently, use $A+A^\top$ as the adjacency matrix of the undirected network.

To quantify similarity between two partitions of the same network, we calculate the Jaccard index for each pair of partitions~\cite{Fortunato2010}. The Jaccard index is defined by
\begin{equation}
J \equiv \frac{a_{1}}{a_{1} + a_{0}},
\end{equation} 
where $a_0$ and $a_1$ are the number of node pairs that are classified to the same community in only one partition and in both partitions, respectively.
If $J=1$, the two partitions are exactly the same. If $J=0$, they completely disagree.

\subsection{Citation networks}\label{sec:citation_net}

\begin{table}
\centering
\caption{Modularity values for the citation networks. The results obtained by maximization of $Q^{\rm und}$,~$Q^{\rm dir}$, and~$Q^{\rm dag}$ using the spectral method are shown in the rows labelled ``s--und'', ``s--dir'', and ``s--dag'', respectively. The results obtained by maximization of $Q^{\rm und}$ using the Louvain method are shown in the rows labeled ``L--und''. $N_{\rm c}$ denotes the number of communities.}
\label{tab:Hep_modularity}
\begin{tabular}{llllll}
\hline\noalign{\smallskip}
Data &	Method	& $N_{\rm c}$ & $Q^{\rm und}$ & $Q^{\rm dir}$ & $Q^{\rm dag}$\\
\hline\noalign{\smallskip}
HepPh &s--und 	&	19	& 	0.7292	&	0.7292 	& 	0.7259 \\
&s--dir 	&	18	& 	0.7288	& 	0.7288	&	0.7257 \\
&s--dag 	&	16	&	0.7300	&	0.7300	& 	0.7263 \\
&L--und 	&	19	&	\vect{0.7331}	&	\vect{0.7332}	& 	\vect{0.7292} \\
\noalign{\smallskip}
\noalign{\smallskip}\hline\noalign{\smallskip}
HepTh &s--und 	&	21	& 	\vect{0.6569}	&	\vect{0.6571} 	& 	\vect{0.6332}\\
&s--dir 	&	30	& 	0.6447	& 	0.6450	&	0.6207 \\
&s--dag	 	&	20	&	0.6452	&	0.6453	& 	0.6320 \\
&L--und 	&	28	&	0.6558	&	0.6561	& 	0.6266 \\
\noalign{\smallskip}\hline
\end{tabular}
\vspace{10pt}
\centering
\caption{Jaccard indices between the partitions obtained for the citation networks.}
\label{tab:Hep_jaccard}
\begin{tabular}{llcccc}
\hline\noalign{\smallskip}
Data &		Method		&	s--und	& s--dir & s--dag & L--und \\
\noalign{\smallskip}
\hline\noalign{\smallskip}
HepPh &s--und	&	1		& 	0.6383  &	0.4116	&	0.4287	\\
&s--dir 		&	0.6383	& 		1	&	0.4298	& 	0.3905	\\
&s--dag 		&	0.4116	&	0.4298	&		1	& 	0.4557	\\
&L--und 		&	0.4287	&	0.3905	&	0.4557	& 		1	\\
\noalign{\smallskip}
\hline\noalign{\smallskip}
HepTh &s--und	&	1		& 	0.3154	&	0.2918	&	0.2859	\\
&s--dir 		&	0.3154	& 		1	&	0.3417	& 	0.3262	\\
&s--dag 		&	0.2918	&	0.3417	&		1	& 	0.2804	\\
&L--und 		&	0.2859	&	0.3262	&	0.2804 & 		1	\\
\noalign{\smallskip}\hline
\end{tabular}

\vspace{10pt}
\centering
\caption{
Modularity values for the \textit{E.~coli} transcriptional regulation network.
See the caption of Table~\ref{tab:Hep_modularity} for legends.}
\label{tab:ecoli_modularity}
\begin{tabular}{llllll}
\hline\noalign{\smallskip}
Data &		Method	& $N_{\rm c}$ & $Q^{\rm und}$ & $Q^{\rm dir}$ & $Q^{\rm dag}$\\
\noalign{\smallskip}\hline\noalign{\smallskip}
\textit{E.~coli} &s--und 	&	13	& 	\vect{0.7366}	&	\vect{0.7373}	& 	\vect{0.6998} \\
&s--dir 		&	13	& 	0.7325	& 	0.7332	&	0.6938 \\
&s--dag	 		&	13	&	0.7136	&	0.7142	& 	0.6988 \\
&L--und 		&	13	&	0.7357	&	0.7364	& 	0.6972 \\
\noalign{\smallskip}\hline
\end{tabular}
\vspace{10pt}
\centering
\caption{
Jaccard indices between the partitions obtained for the \textit{E.~coli} transcriptional regulation network.}
\label{tab:ecoli_jaccard}
\begin{tabular}{llcccc}
\hline\noalign{\smallskip}
Data &		Method		&	s--und	& s--dir & s--dag & L--und \\
\noalign{\smallskip}
\hline\noalign{\smallskip}
\textit{E.~coli} &s--und	&	1			&	 0.7885	&	0.6747		&	0.8875	\\
&s--dir 		&	0.7885		& 		1		&	0.5998		& 	0.7927	\\
&s--dag 		&	0.6747		&	 0.5998	&		1	&	0.6699	\\
&L--und 		&	0.8875		&	 0.7927	&	0.6699		& 	1		\\
\noalign{\smallskip}\hline
\end{tabular}
\end{table}

As an example of temporal networks represented by DAGs, we study two citation networks of articles posted on the e-print archive arXiv.org. We use data on the High Energy Physics Phenomenology (HepPh) and  High Energy Physics Theory (HepTh) sections of arXiv, which exhaustively cover the citations between January 1993 and April 2003 \cite{Leskovec2005,Gehrke2003,Hep-url}.

The HepPh and HepTh data contain 34,546 and 27,770 nodes, and 421,578 and 352,807 links, respectively.
There are four mutually exclusive types of nodes and links excluded from the following analysis.
First, we discard 44 and 39 self-loops in the HepPh and HepTh networks, respectively.
Second, we discard 2,622 and 1,475 links in HepPh and HepTh networks, respectively, which contradict the arrow of time (i.e., articles citing others newer than themselves). Third, in the publicly available data of the HepPh network, the information about the publication date is not provided for the articles posted after 11 March 2002.
Because we cannot assign a date-based layer $\ell_i$ to these articles, we remove 3,985 such articles. Fourth, we remove 74,246 links in the HepPh network that are incident to at least one node whose date information is missing. 
Although we have removed a fraction $74,276/421,578=0.176$ of the links according to the fourth criterion, all removed links are those incident to the newest nodes. Therefore, the network restricted to the dates before 11 March 2002, which we effectively consider in the following, is not affected by the removal of these links.

After removing self-loops, links contradicting the arrow of time, nodes without the information about the publication date, and the links incident to these nodes, we obtain 30,561 nodes and 344,666 links in the HepPh network and 27,770 nodes and 351,293 links in the HepTh network.
We use the largest weakly connected component of each network. Layers are defined by the date of publication. It should be noted that some layers had no node because no article was published on that date. This fact does not change the following results because $Q^{\rm dag}$ and its optimization procedure described in the previous sections are not affected by empty layers.
Some basic quantities of the largest weakly connected component of the two networks are summarized in Table~\ref{tab:data}.

Modularity values obtained after the maximization procedure are summarized for the two networks in Table~\ref{tab:Hep_modularity}. A row corresponds to a combination of the maximized modularity measure
(i.e., $Q^{\rm und}$,~$Q^{\rm dir}$, or $Q^{\rm dag}$) and the maximization method (i.e., spectral or Louvain). For example, s--und represents the spectral method for maximizing $Q^{\rm und}$.
The columns correspond to the three modularity values measured at the end of the maximization procedure, given a modularity maximization method corresponding to a row.

The modularity values shown in Table~\ref{tab:Hep_modularity} seem to be large. 
It turns out that, in both networks, a partitioning method designed for $Q^{\rm und}$ (i.e., L--und for the HepPh network and s--und for the HepTh network) has yielded the largest value for not only  $Q^{\rm und}$ but also $Q^{\rm dir}$ and $Q^{\rm dag}$. Therefore, for these networks, it is better to use a method designed for maximizing $Q^{\rm und}$ to (also) maximize $Q^{\rm dag}$ than a method designed for maximizing $Q^{\rm dag}$. Table \ref{tab:Hep_modularity} also indicates that the spectral method designed for maximizing $Q^{\rm dag}$ (s--dag) does not fall far behind s--und or L--und in terms of the obtained $Q^{\rm dag}$ value. In fact, regardless of the method, we obtained similar values of $Q^{\rm und}$, $Q^{\rm dir}$, and $Q^{\rm dag}$.

The similarity between each pair of the four optimized partitions for each network
(Table \ref{tab:Hep_modularity}) is shown in Table~\ref{tab:Hep_jaccard}, where similarity is measured in terms of 
the Jaccard index. We find that the Jaccard indices are not very large, in particular for the HepTh network, although the different partitioning methods have yielded close modularity values, as shown in Table \ref{tab:Hep_modularity}.
Nevertheless, the Jaccard index may not be very large even between partitions with close modularity values
obtained from the same stochastic partitioning algorithm applied to the same network. For example, the Jaccard index values between 0.5 and 0.9 reported in Ref.~\cite{Raghavan2007} are larger than but comparable with the present values. It has also been reported that partitioning results obtained from the same network yielding 
close modularity values can be fairly different \cite{Good2010,Fortunato2010%
%
%
}.

\subsection{Transcriptional regulation network}

\begin{table}
\centering
\caption{Modularity values for five dominance networks. The results obtained with the optimal modularity maximization method are shown in rows labeled ``o--und'', ``o--dir'', and ``o--dag''. In the table, ``--'' indicates that the optimal method has not terminated.
See the caption of Table~\ref{tab:Hep_modularity} for the other legends.}
\label{tab:ant_modularity}
\begin{tabular}{llllll}
\hline\noalign{\smallskip}
Data &		Method	& $N_c$ & $Q^{\rm und}$ & $Q^{\rm dir}$ & $Q^{\rm dag}$\\
\noalign{\smallskip}\hline\noalign{\smallskip}
C1&s--und 		&	4	& 	0.2990	&	0.3258 	& 	0.3393 \\
&s--dir 		&	3	& 	0.2943	& 	0.3317	&	0.3421 \\
&s--dag 		&	3	&	0.2812	&	0.3282	& 	0.3468 \\
&L--und 		&	4	&	\vect{0.3288}	&	0.3401	& 	0.3633 \\
&o--und 		&	4	&	\vect{0.3288}	&	0.3401	& 	0.3633 \\
&o--dir 		&	4	&	0.3157	&	\vect{0.3484}	& 	\vect{0.3662} \\
&o--dag 		&	4	&	0.3157	&	\vect{0.3484}	& 	\vect{0.3662} \\
\noalign{\smallskip}
\hline\noalign{\smallskip}
C2&s--und 	&	4	& 	0.3522	&	0.3593 	& 	0.3729 \\
&s--dir 	&	4	& 	\vect{0.3607}	& 	0.3702	&	\vect{0.3951} \\
&s--dag 	&	4	&	\vect{0.3607}	&	0.3702 & 	\vect{0.3951} \\
&L--und 	&	4	&	\vect{0.3607}	&	0.3702 	& 	\vect{0.3951} \\
&o--und		&	4	&	\vect{0.3607}	&	0.3702 	& 	\vect{0.3951} \\
&o--dir 	&	5	&	0.3526	&	\vect{0.3716}	& 	0.3890 \\
&o--dag 	&	4	&	\vect{0.3607}	&	0.3702	& 	\vect{0.3951} \\
\noalign{\smallskip}
\hline\noalign{\smallskip}
C3&s--und 	&	5	& 	0.2339	&	0.2372 	& 	0.2503 \\
&s--dir 	&	5	& 	0.2311	& 	\vect{0.2594}	&	0.2643 \\
&s--dag 	&	6	&	\vect{0.2416}	&	0.2531	& 	\vect{0.2654} \\
&L--und 	&	3	&	0.2378	&	 0.2421	& 	0.2572 \\
&o--und 	&	--	&	--			&	--			& --	 \\
&o--dir		&	5	&	0.2311	&	\vect{0.2594}	&   0.2643	\\
&o--dag 	&	6	&	\vect{0.2416	} &	0.2531	&   \vect{0.2654}	 \\
\noalign{\smallskip}
\hline\noalign{\smallskip}
C4&s--und 	&	5	& 	\vect{0.2690}	&	0.2782 	& 	0.2882 \\
&s--dir 	&	6	& 	0.2612	& 	\vect{0.2788}	&	0.2899 \\
&s--dag 	&	5	&	0.2604	&	0.2767	& 	\vect{0.2914} \\
&L--und 	&	5	&	0.2626	&	0.2656	& 	0.2766 \\
&o--und		&	--	&	--	&	--	& --	 \\
&o--dir 	&	--	&	--	&	--	& --	 \\
&o--dag 	&	--	&	--	&	--	& --	 \\
\noalign{\smallskip}
\hline\noalign{\smallskip}
C6&s--und 	&	6	& 	0.3374	&	0.3396 	& 	0.3512 \\
&s--dir 	&	5	& 	0.3308	& 	0.3422	&	0.3412 \\
&s--dag 	&	6	&	0.3348	&	0.3428	& 	0.3600 \\
&L--und 	&	5	&	0.3467	&	0.3504	& 	0.3493 \\
&o--und		&	5	&	\vect{0.3487}	&	\vect{0.3529}	& 	 0.3521 \\
&o--dir 	&	5	&	\vect{0.3487	} &	\vect{0.3529}	& 	 0.3521 \\
&o--dag 	&	6	&	0.3356	&	0.3440	& 	 \vect{0.3606} \\
\noalign{\smallskip}\hline
\end{tabular}
\end{table}

\begin{table*}
\centering
\caption{Jaccard indices between the partitions obtained for the dominance networks.}
\label{tab:ant_jaccard}
\begin{tabular}{llccccccc}
\hline\noalign{\smallskip}
Data &		Method		&	s--und	& s--dir & s--dag & L--und & o--und & o--dir & o--dag\\
\noalign{\smallskip}
\hline\noalign{\smallskip}
C1 &s--und	&	1		& 	0.7288	&	0.6452	&	0.5965	&	0.5965	&	0.5439	&  	0.5439	\\
&s--dir 	&	0.7288	& 		1	&	0.5946	& 	0.4658 &	0.4658	&	0.4857	&	0.4857	\\
&s--dag 	&	0.6452	&	0.5946	&		1	& 	0.5970	&	0.5970	&	0.7049	&  	0.7049	\\
&L--und 	&	0.5965	&	0.4658	&	0.5970	& 		1	&	1		&	0.8235	&  	0.8235	\\
&o--und		&	0.5965	&	0.4658	&	0.5970	&	1		&	1	&	0.8235	&  	0.8235	\\
&o--dir		&	0.5439	&	0.4857	&	0.7049	&	0.8235	&	0.8235	&		1	&  	1		\\
&o--dag		&	0.5439	&	0.4857	&	0.7049	&	0.8235	&	0.8235	&		1	&  	1		\\
\noalign{\smallskip}
\hline\noalign{\smallskip}
C2 &s--und		&	1		& 	0.7083	&	0.7083	&	0.7083	&	0.7083	&	0.5753	&  	0.7083	\\
&s--dir 		&	0.7083	& 		1	&	1		& 	1		&	1		&	0.7724	&	1		\\
&s--dag 		&	0.7083	&	1		&		1	& 		1	&	1		&	0.7724 &		1	\\
&L--und 		&	0.7083	&	1		&		1	& 		1	&	1		&	0.7724	&  		1 \\
&o--und			&	0.7083	&	1		&	1		&	1		&	1		&	0.7724	&  	1		\\
&o--dir			&	0.5753	&	0.7724	&	0.7724	&	0.7724	&	0.7724	&		1	&  	0.7724	\\
&o--dag			&	0.7083	&	1		&		1	&	1		&	1		&	0.7724	&  		1	\\
\noalign{\smallskip}
\hline\noalign{\smallskip}
C3 &s--und		&	1		& 	0.2727	&	0.2755	&	0.3028	&	--		&	0.2727	&  	0.2755	\\
&s--dir 		&	0.2727	& 		1	&	0.6687	& 	0.3275	&	--		&		 1	&	0.6687	\\
&s--dag 		&	0.2755 &	0.6687	&		1	& 	0.3548	&	--		&	0.6687 	&		 1	\\
&L--und 		&	0.3028	&	0.3275	&	0.3548	& 		1	&	--		&	0.3275	&  	0.3548	\\
&o--und			&	--		&	--		&	--		&	--		&		1	&		--	&  		--	\\
&o--dir			&	0.2727	&	1		&	0.6687	&	0.3275	&	--		&		1	& 	0.6687	\\
&o--dag			&	0.2755	&	0.6687	&	1		&	0.3548	&	--		&	0.6687	&  		1	\\
\noalign{\smallskip}
\hline\noalign{\smallskip}
C4 &s--und		&	1		& 	0.5726	&	0.5928	&	0.4807	&	--		&		--	&  		--	\\
&s--dir 		&	0.5726	& 		1	&	0.8047	& 	0.6039	&	--		&		--	&		--	\\
&s--dag 		&	0.5928	&	0.8047	&		1	& 	0.5827	&	--		&		-- 	&		--	\\
&L--und 		&	0.4807	&	0.6039	&	0.5827	& 		1	&	--		&		--	&  		--	\\
&o--und			&	--		&	--		&	--		&	--		&		1	&		--	&  		--	\\
&o--dir			&	--		&	--		&	--		&	--		&	--		&		1	& 	 	--	\\
&o--dag			&	--		&	--		&	--		&	--		&	--		&		--	&  		1	\\
\noalign{\smallskip}
\hline\noalign{\smallskip}
C6 &s--und		&	1		& 	0.5653	&	0.4153	&	0.5841	&	0.5697	&	0.5697	&  	0.3254	\\
&s--dir 		&	0.5653	& 		1	&	0.3055	& 	0.5779	&	0.6641	&	0.6641	&	0.2852	\\
&s--dag 		&	0.4153	&	0.3055	&		1	& 	0.3699	&	0.3121	&	0.3121 	&	0.7118	\\
&L--und 		&	0.5841	&	0.5779	&	0.3699	& 		1	&	0.7588	&	0.7588 	&  	0.2959	\\
&o--und			&	0.5697	&	0.6641	&	0.3121	&	0.7588	&		1	&		1	&  	0.3448	\\
&o--dir			&	0.5697	&	0.6641	&	0.3121	&	0.7588	&		1	&		1	& 	0.3448	\\
&o--dag			&	0.3254	&	0.2852	&	0.7118	&	0.2959	&	0.3448	&	0.3448	&  		1	\\
\noalign{\smallskip}\hline
\end{tabular}
\end{table*}

We study the compartmentalization of the transcriptional regulation network of the bacteria \textit{Escherichia coli}~(\textit{E.~coli}) using publicly available data
\cite{Mangan2003,Ecoli-url}.
Nodes of this network are operons. Links are directed from the operon that encodes a transcription factor to an operon regulated by the transcription factor. The network, which is a DAG, contains 423 nodes and 519 directed links without self loops. We extracted the largest weakly connected component, which consisted of 328 nodes and 456 links. Layers were determined by a 
leaf-removal algorithm~\cite{Eades1993}. In this algorithm, the nodes with out-degree zero are classified to the lowest layer. Then, these nodes together with links connecting to these nodes are removed. The nodes with out-degree zero in the remaining network are classified to next lowest layer. This procedure is iterated until all nodes are exhausted. The basic quantities of this network are summarized in Table~\ref{tab:data}.

The modularity values obtained from the four modularity maximization algorithms are shown in Table~\ref{tab:ecoli_modularity}.
Algorithm s--und produces the largest $Q^{\rm und}$, $Q^{\rm dir}$, and $Q^{\rm dag}$, similarly to the results for the HepTh network (Table~\ref{tab:Hep_modularity}).
The Jaccard indices are shown in Table~\ref{tab:ecoli_jaccard}. They are much larger than those for the citation networks (Table~\ref{tab:Hep_jaccard}), suggesting that the partitions
%
%
obtained by the different methods are relatively similar to each other for the present network.

\subsection{Dominance networks}

As a final example, we examine DAGs induced by dominance hierarchies in ant colonies using the data in Ref.~\cite{Shimoji2014}. The data set contains aggression-based hierarchy among workers in six ant colonies of species \textit{Diacamma sp}. Nodes represent individual female ant workers and a link is drawn from the attacking to the attacked ants, where an attack is defined as a bite and jerk. Layers are determined by the leaf-removal algorithm used for the transcriptional regulation network.
The data set contains six colonies, five of whose largest weakly connected components are DAGs that we use in the following analysis.
Basic quantities of the largest weakly connected component of the five DAGs are summarized in Table~\ref{tab:data}. 

The present DAGs are smaller than those analyzed in the previous sections. Therefore, for some networks, we were able to calculate the partition
that exactly maximized the modularity using integer linear programming~\cite{Brandes2008}. In this method, originally applied to $Q^{\rm und}$, variables $X_{ij} \in \{0,1\}$ are defined for every pair of nodes $i$ and $j$, where $X_{ij}=1$ ($X_{ij}=0$) indicates that $i$ and $j$ are classified to the same community (different communities). Then, we rewrite the modularity in terms of $X_{ij}$ by replacing $\delta_{c_ic_j}$ by $X_{ij}$ in Eqs.~\eqref{eq:Q_undirected},~\eqref{eq:Q_directed}, and~\eqref{eq:Q_ordered} for $Q^{\rm und}$, $Q^{\rm dir}$, and $Q^{\rm dag}$, respectively. The variables need to satisfy $X_{ii}=1$, $X_{ij} = X_{ji}$, and $
X_{ij} + X_{jh} - 2 X_{ih} \le 1$ for all $1\le h,i,j\le N$, which defines an integer linear program for exact maximization of modularity.

The results for modularity maximization and the Jaccard indices for the obtained partitions are shown in 
Tables~\ref{tab:ant_modularity} and \ref{tab:ant_jaccard}, respectively.
Table~\ref{tab:ant_modularity} indicates that the obtained modularity values vary across colonies (i.e., networks) and do not solely depend on the size of the network. Therefore, different colonies may have different degrees of community structure.
The table also indicates that s--dag realizes the largest $Q^{\rm dag}$ values in all colonies except C1, among the four heuristic methods (i.e., s--und, s--dir, s--dag, and L--und). Algorithm s--dag realizes the optimal $Q^{\rm dag}$ value obtained by o--dag for C2 and C3. The $Q^{\rm dag}$ value obtained by s--dag is also close to that obtained by o--dag for C6, whereas this is not the case for C1. Similarity among the partitions obtained by the different algorithms depends on colonies without clear patterns (Table~\ref{tab:ant_jaccard}).

\section{Conclusions}

We have proposed a modularity measure for DAGs and a spectral algorithm to maximize it by extending the spectral algorithm developed for undirected networks. We found that the obtained modularity values are rather independent of whether we used the proposed algorithm or other algorithms known for less restricted networks such as undirected or directed networks allowing for cycles. Therefore, up to our numerical efforts, simply applying modularity maximization methods for undirected networks to DAGs may be practically innocuous. We stress that we have reached this conclusion by actually developing a modularity measure for DAGs and testing it against previous methods using several data sets.

The spectral method was presented as an example heuristic for maximizing the proposed modularity measure. The proposed modularity measure can be also maximized by other approximate methods, which may surpass the performance of the spectral method.
 
We acknowledge Steve Gregory and Kohei Tamura for discussion and careful reading of the manuscript. L.S. acknowledges the support provided through DAAD. N.M. acknowledges the support provided through JST, CREST, and JST, ERATO, Kawarabayashi Large Graph Project.

T.T. and N.M. designed the research. L.S., T.T., and N.M. developed the theory. L.S. and T.T. analyzed the data. L.S., T.T., and N.M. discussed the results and wrote the paper.

\appendix

\section{Fine tuning after bipartitioning}\label{app:fine tuning}

In the spectral methods for maximizing $Q^{\rm und}$, $Q^{\rm dir}$, and $Q^{\rm dag}$, we carried out the following fine tuning procedure every time after a community was bipartitioned.
Suppose that we have bipartitioned a community $C$ into two communities~$C_1$ and~$C_2$.
For every node in~$C$, we tentatively moved the node to the opposite community and calculated the change in the targeted modularity value. Then, we adopted the attempted move that maximized the modularity under the condition that the modularity increased by the move. We repeated this procedure under the constraint that each node was moved at most once, until no further increase in modularity was possible. This fine tuning procedure slightly modifies that proposed for $Q^{\rm und}$ \cite{Newman2006,Newman2006a}. We modified the original procedure because it involved repetition of a procedure similar to the aforementioned one and did not terminate on our data when we attempted to maximize $Q^{\rm dag}$.

\section{Postprocessing}\label{app:postprocessing}

After the completion of modularity maximization with the spectral or Louvain method, we carried out the following 
postprocessing algorithm to enhance the targeted modularity value. Our postprocessing algorithm resembles that employed in other modularity maximization algorithms~\cite{Blondel2008,Newman2004a%
%
%
}.

Step 1: For a given node $i$ ($1\le i\le N$), we tentatively moved it to the community to which each adjacent node of $i$ belonged. The move attempt that increased the modularity by the largest amount was adopted.
We scanned the $N$ nodes in random order. We neglected the direction of link when judging adjacency in the case of directed networks including DAGs.

Step 2: First, we selected the smallest community in terms of the number of nodes, called the seed community. Second, we tentatively merged the seed community with each of the remaining communities to which the seed community is directly connected by a link regardless of the direction. Third, we measured the modularity value. We adopted the merge attempt that yielded the largest modularity but only when the modularity increased by the merge.
Then, among the community that had never been selected as seed community, we selected the smallest community as seed community and attempted merger. We repeated the same procedure until no further increase in the modularity occurred.

We applied step 1 and then step 2. We repeated the combination of the two steps ten times. We also verified that 
swapping the order of the two steps had little impacts on the final modularity value.

%

\end{document}